\documentclass[journal]{IEEEtran}
\def\baselinestretch{0.94}
\usepackage{amsmath}
\usepackage{amssymb,xspace,subfigure,amsfonts}
\usepackage{epsfig,epsf,epstopdf,graphicx}
\usepackage[ruled]{algorithm2e}
\usepackage{lineno}
\usepackage{mathtools}
\usepackage{cite}
\usepackage{float}
\usepackage{color}
\begin{document}
\title{Multivariate Copula Spatial Dependency in One Bit Compressed Sensing}

\author{Zahra~Sadeghigol,
Hadi~Zayyani, Hamidreza~Abin,
        and~Farrokh~Marvasti,~\IEEEmembership{Senior Member,~IEEE}
\thanks{Z. Sadeghigol is with the Department
of Electrical Engineering, Sharif University of Technology, Tehran, Iran (e-mail: sadeghigol@gmail.com).}
\thanks{H. Zayyani is with the faculty of Electrical and Computer Engineering, Qom University of Technology (QUT), Qom, Iran (e-mail: zayyani@qut.ac.ir).}
\thanks{H. Abin and F. Marvasti are with the Department
of Electrical Engineering, Sharif University of Technology and Advance Communication Research Institute (ACRI), Tehran, Iran (e-mails: Hamidreza.abin@gmail.com; marvasti@sharif.edu).}
\thanks{This work is fully supported by Iran National Science Foundation (INSF). }}

%

\maketitle

\begin{abstract}
In this letter, the problem of sparse signal reconstruction from one bit compressed sensing measurements is investigated. To solve the problem, a variational Bayes framework with a new statistical multivariate model is used. The dependency of the wavelet decomposition coefficients is modeled with a multivariate Gaussian copula. This model can separate marginal structure of coefficients from their intra scale dependency. In particular, the drawable Gaussian vine copula multivariate double Lomax model is suggested. The reconstructed signal is derived by variational Bayes algorithm which can calculate closed forms for posterior of all unknown parameters and sparse signal. Numerical results illustrate the effectiveness of the proposed model and algorithm compared with the competing approaches in the literature.
\end{abstract}

\begin{IEEEkeywords}
one bit compressed sensing, vine copula model, variational Bayes.
\end{IEEEkeywords}

\IEEEpeerreviewmaketitle

\section{Introduction}
\IEEEPARstart{T}{he} extreme case of quantized Compressed Sensing (CS) \cite{zymnis2010compressed, moshtaghpour2016consistent,zamani2016iterative} which is One bit Compressed Sensing (1b-CS) has enticed consideration newly \cite{boufounos20081, boufounos2009greedy, laska2011trust, jacques2013robust,yan2012robust,plan2013robust,candes2006near,donoho2006compressed,li2015robust, dong2015map, chen2015dictionary, zayyani2016dictionary}. This quantization procedure can achieve greatly cost-effective impact on the hardware. Traditional CS theory can reconstruct a sparse signal from much smaller number of linear measurements than the Nyquist rate \cite{candes2006near},~\cite{donoho2006compressed}. According to the CS paradigm, the procedure of quantization is omitted and assumed that the measurements are real. However, in the quantized CS, some discrete levels are assigned to the measurements. In 1b-CS framework, only two levels are used to represent the measurements \cite{boufounos20081}-\cite{zayyani2016dictionary}. It is proved that only utilizing the sign of the measurements is sufficient to accurately reconstruct the sparse signal \cite{zamani2016iterative}.

There are numerous algorithms suggested for the signal recovery in 1b-CS framework \cite{zamani2016iterative}. In \cite{boufounos20081}, an $\ell_1$-norm minimization is proposed which is known as Renormalized Fixed-Point Iteration (RFPI) algorithm. Authors in \cite{boufounos2009greedy} presented Matching Sign Pursuit (MSP) algorithm for solving the problem. In \cite{jacques2013robust}, a Binary Iterative Hard Thresholding (BIHT) algorithm is appeared which is reported due to better accuracy than MSP. Furthermore, \cite{laska2011trust} introduced a Restricted-Step Shrinkage (RSS) method which is proved to have a guaranteed convergence. In \cite{yan2012robust}, authors developed an Adaptive Outlier Pursuit (AOP) algorithm in the noisy 1b-CS scenario in which sign flip errors may exist. Also, a convex optimization solution is introduced in \cite{plan2013robust} to solve the problem. Moreover, in \cite{li2015robust} a Variational Bayes (VB) algorithm is used for 1b-CS, while in \cite{dong2015map} a Maximum A Posteriori (MAP) approach is presented for the signal recovery. In addition, a dictionary learning-based blind 1b-CS algorithm is suggested in \cite{zayyani2016dictionary}. Recently, authors in \cite{sun2017training} revealed a new training-free 1b-CS approach for wireless neural recording.

All the above algorithms are based on univariate models causing to simple solutions. Nevertheless, these univariate models are not capable to describe statistical manner of wavelet coefficients completely. These simple models ignore an important stochastic property of wavelet coefficients which is the intrascale dependency across the same subband. To overcome this problem, in this paper, a multivariate model is proposed  which is based on copula distribution. In the best of our knowledge, multivariate models have not been considered in 1b-CS, yet. However, in the field of image processing, some researchers have tried to model the joint dependency of wavelet coefficients. Multivariate Gaussian distribution \cite{tzagkarakis2006rotation}, Multivariate Generalized Gaussian Distribution (MGGD) \cite{cho2005multivariate} and Elliptically Contoured Distribution \cite{tan2007multivariate} have been proposed so far. The non-Gaussian joint stochastic manner of wavelet coefficients can be demonstrated by Gaussian Scale Mixture (GSM) model \cite{wainwright2000scale}, \cite{portilla2003image} and \cite{boubchir2010multivariate}.

Some studies have recently worked on copulas for modeling multivariate wavelet coefficients \cite{lasmar2014gaussian,kwitt2011efficient,stitou2009copulas}. None of them considered the 1b-CS problem. The major advantage of copula is its flexibility for choosing various kinds of marginal distributions based on the joint model. Accordingly, our contribution is as follows:

    \begin{enumerate}
      \item A new multivariate model which is named Drawable Gaussian Vine Copula-based Multivariate Double Lomax (DGVC-MDL) is proposed for capturing the spatial intrascale dependencies of wavelet coefficients.
      \item Based on this new proposed model, the full posteriors of unknown variables using VB are derived in closed form.
    \end{enumerate}

Simulation results demonstrate that the proposed DGVC-MDL improves recovery performance in comparison to state of the art methods in the 1b-CS literature.

The outline of the paper is as follows. Section ~\ref{sec: problem} introduces the problem formulation. Then, the multivariate model is presented in Section~\ref{sec: moodel}. The proposed variational inference procedure is illustrated in ~\ref{Variational Inference}. Simulation results are presented in Section~\ref{sec: Sim}. Finally, conclusions are drawn in Section~\ref{sec: con}.

\section{Problem Formulation}
\label{sec: problem}
The problem of sparse signal recovery from 1b-CS measurements can be formulated as follows:
\begin{equation}\label{1}
\textbf{\textit{t}}=\mathrm{sign}\left(\textbf{\textit{y}}\right)
=\mathrm{sign}(\textbf{\textit{Ax}}+\textbf{\textit{w}}),
\end{equation}
where $\textbf{\textit{A}} \in \mathbb{R}^{n \times m}$ is the
measurement matrix, $\textbf{\textit{t}} \in
\mathbb{R}^{n}$ is the sign measurement vector of
$\textbf{\textit{y}} \in \mathbb{R}^{n}$, and $\textbf{\textit{n}}
\in \mathbb{R}^{n}$ is the noise measurement vector which is
assumed to be i.i.d random Gaussian with zero mean and variance $\sigma_n^2$.
Our aim is to estimate the sparse signal $\textbf{\textit{x}} \in
\mathbb{R}^{m}$ from the sign of measurements $\textbf{\textit{t}}$.

\section{Multivariate Statistical Model}
\label{sec: moodel}

\subsection{Basics}

The coefficients intrascale dependencies across subband are extensively considered before \cite{lasmar2014gaussian}. \cite{po2006directional} has studied the amount of intraband and interband coefficients dependency using mutual information. In the image denosing field, \cite{tan2007multivariate} and \cite{portilla2003image} introduced several types of wavelet neighborhood. All these researches mentioned that the overcoming coefficients dependency is based on the intraband spatial structure. Using Chi-plot graphs, \cite{lasmar2014gaussian} demonstrated that the intraband dependency is more important than interorientation and interscale ones. Therefore, in this letter, we assume the intrascale dependency and ignore other kind of dependencies \cite{lasmar2014gaussian}. To capture this dependency, the proposed algorithm introduces DGVC model which can provide an excellent marginal distribution fitting.

Based on wavelet decomposition, scales and orientations subbands are formed. Three kinds of dependencies using DGVC are defined in the proposed algorithm as row dependency, column dependency and diagonal dependency. These three dependencies are modeled by vine copula trees across each row, column and diagonal, respectively.

\begin{figure}
  \centering
  \includegraphics[width=60mm]{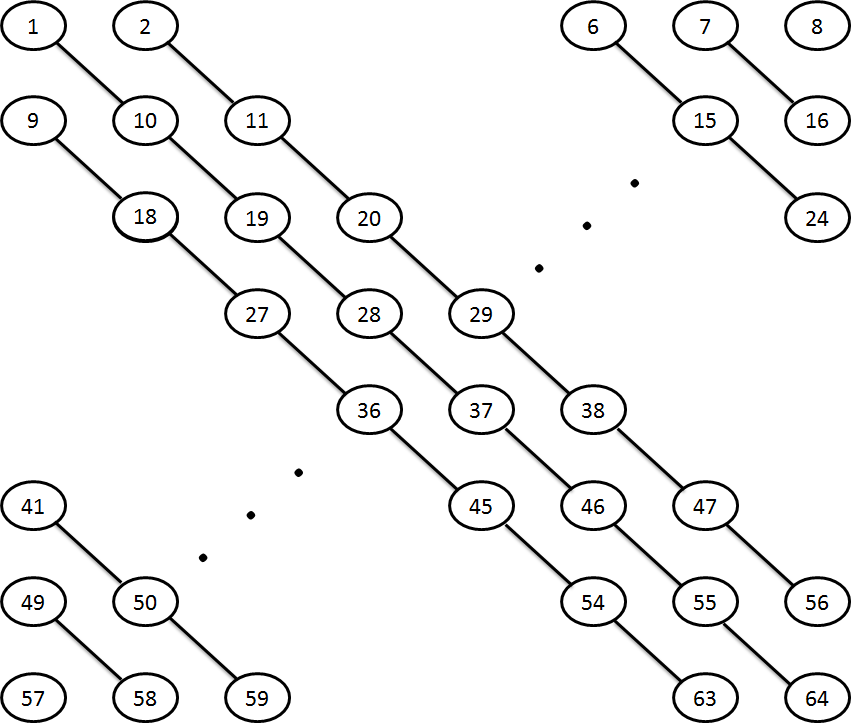}
  \caption{Diagonal dependency in the vine copula root trees.}\label{Fig.1}
\end{figure}

Because of large number of wavelet coefficients, DGVC model is imposed in the proposed DGVC-MDL algorithm to capture the intrascale dependencies in row, column and diagonal directions. The diagonal tree of DGVC is depicted in Fig. \ref{Fig.1}. In this model, each node of the tree has a degree of at most 2, where the degree of a node indicates as the number of connections \cite{brechmann2012truncated}.

\subsection{Multivariate Model Based on DGVC}

In this subsection, the proposed algorithm based on DGVC is introduced.

A copula demonstrates a multivariate distribution with standard uniform marginal distributions \cite{brechmann2012truncated}. The coefficients dependency is completely defined by the copula which is totally independent of the marginal distribution definition.

A copula $C$ can be considered as joint Cumulative Distribution Function (CDF) on ${[0,1]}^d$. Suppose a multivariate random variable $\vec{X}={[{x}_1, ... ,{x}_d]}^T$ with marginal CDFs ${F}_1, ... ,{F}_d$. Sklar's
theorem \cite{sklar1973random} expresses that a unique copula can be found such that:
\begin{equation}\label{2}
  F({x}_1, ... ,{x}_d)=C({F}_1(x_1), ... ,{F}_d(x_d)).
\end{equation}

The joint Probability Density Function (PDF) of $\vec{X}$ can be written as:

\begin{equation}\label{3}
  f({x}_1, ... ,{x}_d)=\left[\prod_{k=1}^{d}f_k(x_k)\right]\times c({F}_1(x_1), ... ,{F}_d(x_d)),
\end{equation}
where $f_k$, $k=1, ...,d$ are the marginal PDFs.

 We consider the Gaussian copula in our proposed algorithm since it completely fits the statistical wavelet coefficient features. Besides the parameters of multivariate model base on Gaussian copula can  be quickly estimated in the closed form by Maximum Likelihood (ML) method. The Gaussian copula density can be derived from (\ref{2}) as follows:

\begin{equation}\label{4}
  c({u}_1, ... ,{u}_d)=\frac{1}{|\Sigma|^{1/2}}\exp\frac{-\vec{v}(|\Sigma|^{-1}-I)\vec{v}}{2},
\end{equation}
where ${u}_i$ is uniform on $[0, 1]$, $\vec{v}=({v}_1, ... ,{v}_d)$ is a vector of transformed observations as ${v}_i=\phi^{-1}({u}_i)$, and $\phi$ stands for the standard Gaussian CDF. $I$ denotes an identity matrix with $d\times d$ dimension and $\Sigma$ is the covariance matrix of $\vec{v}$.

Multivariate copulas in higher dimensions are not proper and have some restrictions. Therefore, a flexible model of copula is essential for capturing the high dimensional dependencies. \cite{bedford2002vines} introduced regular vine copula and \cite{kurowicka2006uncertainty} presented it in detail. We utilize DGVC in the proposed algorithm which its density is given by:

  \begin{equation}\label{5}
\begin{multlined}
    f({x}_1, ... ,{x}_d)=\left[\prod_{k=1}^{d}f_k(x_k)\right] \times \left[\prod_{j=1}^{d-1} \prod_{i=1}^{d-j} c_{i,i+j|i+1,...,i+j-1} \right],
\end{multlined}
\end{equation}
where
\begin{equation}\label{6}
\begin{multlined}
        c_{i,j|i_{1},...,i_{k}}:=c_{i,j|i_{1},...,i_{k}}(F({x_{i}|x_{i_{1}},...,x_{i_{k}}}),F({x_{j}|x_{i_{1}},...,x_{i_{k}}})),
\end{multlined}
\end{equation}
$c_{j(e),k(e)|D(e)}$ is a bivariate copula density in a DGVC tree with the node set $N:={{N}_1, ... ,{N}_{d-1}}$ and the
edge set $E:={{E}_1, ... ,{E}_{d-1}}$. Each edge $e=j(e),k(e)|D(e)$ in ${E}_i$ represents by $c_{j(e),k(e)|D(e)}$. $j(e)$ and $k(e)$ are nominated as the conditioned nodes and $D(e)$ is called as the conditioning set \cite{kurowicka2006uncertainty}. More details in this regard can be found  in \cite{kurowicka2006uncertainty}.

\subsection{Multivariate DGVC Modeling Estimation}
Suppose that $\boldsymbol{\zeta}=(\zeta_{1},...,\zeta_{d})$ and $\Sigma$ represent the marginal PDF parameters  and the copula covariance matrix, respectively. Multivariate GVC hyperparameters are $\boldsymbol{\theta}=(\boldsymbol{\zeta},\Sigma)$ which should be estimated.

\cite{joe1996estimation} represented that the copula covariance matrix $\hat{\Sigma}$ can be evaluated separately from the marginal PDF parameters $\hat{\boldsymbol{\zeta}}=(\hat{\zeta_{1}},...,\hat{\zeta_{d}})$. This assumption simplifies the estimation mechanism of hyperparameters as follows:

1) $\hat{\boldsymbol{\zeta}}=(\hat{\zeta_{1}},...,\hat{\zeta_{d}})$ can be estimated using VB inference which will be described in the next section.

2) Utilizing the ML estimator, $\hat{\Sigma}$ can be evaluated in the following procedure.

 \textbf{ML estimator for} $\boldsymbol{\hat{\Sigma}}$:
At each scale of wavelet decomposition, all coefficients $x$ should be transformed to $v=\phi^{-1}(F(x|\zeta))$. The transformed coefficients are restructured into a matrix $H=[\vec{V}_{1},...,\vec{V}_{M_{s}}]$ and $M_{s}$ is the number of wavelet coefficients in scale $s$. $\vec{V_{i}}$ is a vector which contains a reference coefficient and its neighbors. It can be denoted as $\vec{V}_{i}=[v_{i,1} ... v_{i,L}]^{T}$ and $L$ is the neighborhood size. Therefore, $\boldsymbol{\Sigma}$ can be estimated using ML algorithm which leads to the sample covariance
matrix of $\vec{V}_{1},...,\vec{V}_{M_{s}}$:

\begin{equation}\label{7}
\begin{multlined}
       \boldsymbol{\hat{\Sigma}}=\frac{1}{M_{s}}\sum_{i=1}^{M_{s}}{\vec{V}_{i}}{\vec{V}_{i}^{T}}=\frac{1}{M_{s}}\boldsymbol{H}\boldsymbol{H}^{T}.
\end{multlined}
\end{equation}

One of the rich distribution to represent the marginal manner of wavelet coefficients is Double Lomax (DL) PDF which is used in our proposed DGVC-MDL model.

\textbf{ML estimator for} $\boldsymbol{\hat{\eta}}$ and $\boldsymbol{\hat{f}}$:
To evaluate $\boldsymbol{H}$ in (\ref{7}), parameters of DL distribution should be computed which is given by:

\begin{equation}\label{8}
  \left\{\boldsymbol{\hat{\eta}}, \boldsymbol{\hat{f}}\right\}=\underset{\eta, f}{\text{argmax log}} \prod_{i=1}^{M_s}{f(\overrightarrow{x_i}|\eta,f)},
\end{equation}
where $f$ is DL PDF as following\cite{gu2013variational}:
\begin{equation}\label{9}
  f(x|\eta, f)=\frac{\eta}{2}\left( 1+\frac{\eta|x|}{f}\right)^{-(f+1)},
\end{equation}
 with  $\eta>0$ as the scale parameter and $f>0$ as the shape parameter \cite{kleiber2003statistical}.
$\boldsymbol{\hat{\eta}}$ can be estimated by:
\begin{equation}\label{10}
\begin{multlined}
  \boldsymbol{\hat{\eta}}=\frac{M_{s}}{\sum_{i=1}^{M_{s}}(\frac{(\boldsymbol{\hat{f}}+1)|x_{i}|}{\boldsymbol{\hat{f}}+\boldsymbol{\hat{\eta}}|x_{i}|})},
\end{multlined}
\end{equation}
$\boldsymbol{\hat{f}}$ can be evaluated using   the  numerical Newton-Raphson iterative procedure.

\subsection{Multivariate Double Lomax Model Based on Gaussian Vine Copula}

The proposed DGVC-MDL model is defined by

\begin{equation}\label{11}
\begin{multlined}
       f_{\text{DGVC-MDL}}(\boldsymbol{\vec{x}}|\boldsymbol{\theta})= \left(\frac{\eta}{2}\right)^{d}\left(1+\frac{\eta|x|}{f}\right)^{-d(f+1)}\\
       \times \left[\prod_{j=1}^{d-1} \prod_{i=1}^{d-j} c_{i,i+j|i+1,...,i+j-1} \right].
\end{multlined}
\end{equation}
The hyperparameters set is $\boldsymbol{\theta}=\left(\eta, f, \boldsymbol{\Sigma}\right)$ and $d$ is the neighborhood size. The covariance matrix of $\vec{V}$ is $\boldsymbol{\Sigma}$ using $V_{i}=\phi^{-1}\left(F(x_{i}|\eta,f)\right)$ which $\phi$ denotes the standard normal CDF and $F(t|\eta,f)=\frac{1}{2}\left(\text{sgn}(t)+1\right)\left(1-\frac{1}{2}\left(\frac{\eta t}{f}+1\right)^{-f}\right)-\frac{1}{4}\left(\text{sgn}(t)-1\right)\left(1-\frac{\eta t}{f}\right)^{-f}$ is the CDF of DL distribution.


In order to impose VB on the proposed DGVC-MDL probabilistic model in (\ref{11}), we have to consider the hierarchical form of DL distribution which is introduced by:
\begin{equation}\label{12}
\begin{multlined}
  p(\boldsymbol{x}|\tau)=\prod_{i=1}^{n}p(x_{i}|\tau_{i})=\prod_{i=1}^{n}\mathcal{N}(x_{i}|0,\tau_{i}), \\
\end{multlined}
\end{equation}
\begin{equation}\label{13}
  p(\boldsymbol{\tau}|\lambda)=\prod_{i=1}^{n}\mathcal{E}xp(\frac{\tau_{i}|\lambda^{2}}{2}),\\
\end{equation}
\begin{equation}\label{14}
  p(\boldsymbol{\lambda})=\prod_{i=1}^{n}\mathcal{G}amma(\lambda_{i}|f,f),
\end{equation}
where $\tau$ and $\lambda$  are the precision of Gaussian PDF and exponential PDF parameter, respectively.

%

\section{Proposed Variational Inference Procedure}
\label{Variational Inference}
To derive the posterior PDF of unknown variables, VB inference is
imposed in the proposed algorithm. Suppose
$\Psi\triangleq\left\{x,\tau,\lambda\right\}$ is assigned to all
variables in our proposed model. To model the sign function in (\ref{1}), we use the derivation of \cite{li2015robust}.

An approximation of the posterior PDF
$p(\boldsymbol{\Psi}|\boldsymbol{t})$ can be obtained by maximizing
\begin{equation}\label{16}
  L(q)=\int q(\boldsymbol{\Psi})\ln \frac{p(\boldsymbol{t},
  \boldsymbol{\Psi})}{q(\boldsymbol{\Psi})}d\boldsymbol{\Psi},
\end{equation}
where $p(\boldsymbol{t}, \boldsymbol{\Psi})$ is stated as

\begin{equation}\label{17}
  p(\boldsymbol{t}, \boldsymbol{\Psi})=p(\boldsymbol{t}|\boldsymbol{x})p(\boldsymbol{x}|\tau)p(\boldsymbol{\tau}|\lambda)p(\textbf{copula}),
\end{equation}
$p(\textbf{copula})$ is the second part of (\ref{5}).
$p(\boldsymbol{t}|\boldsymbol{x})$ in (\ref{17}) is difficult to evaluate. To overcome this problem, \cite{li2015robust} found a lower bound on $L(q)$ using the following inequality

\begin{equation}\label{18}
\begin{multlined}
  \sigma(y)^{t}[1-\sigma(y)]^{1-t}=\sigma(z)\\
  \geq  \sigma(\delta) exp\left(\frac{z-\delta}{2}-\lambda(\delta)(z^{2}-\delta^{2})\right),
 \end{multlined}
\end{equation}
where $z=(2t-1)y$, $\lambda(\delta)=(1/4\delta)\tanh \delta/2)$, $\tanh(x)$ is stated to the hyperbolic tangent function which is $\tanh(x)\triangleq \left(\exp(x)-\exp(-x)\right)/\left(\exp(x)+\exp(-x)\right)$. The equality is achieved when $\delta=z$. Utilizing equation (\ref{18}), \cite{li2015robust} obtained

 \begin{equation}\label{19}
\begin{multlined}
  p(\boldsymbol{t}|\boldsymbol{x})\geq F(\boldsymbol{t}, \boldsymbol{x}, \boldsymbol{\delta})\\
  \triangleq \prod_{i=1}^{n} \sigma(\delta_{i}) \exp\left(\frac{z_{i}-\delta_{i}}{2}-\lambda(\delta_{i})(z_{i}^{2}-\delta_{i}^{2})\right).
 \end{multlined}
\end{equation}
 Then \cite{li2015robust} defined $G(\boldsymbol{t}, \boldsymbol{\Psi}, \boldsymbol{\delta})\triangleq F(\boldsymbol{t}, \boldsymbol{x},\boldsymbol{\delta}) p(\boldsymbol{x}|\tau) p(\boldsymbol{\tau}|\lambda)p(\textbf{copula})$ to compute a lower bound on $L(q)$. The different steps of VB procedure are represented as follows.

 \textbf{1. Update of $q_{x}(x)$}: The variational approximation of the posterior PDF $q_{x}(x)$ can be derived by

\begin{eqnarray}\label{20}
  \ln q_{x}(\boldsymbol{x}) &\propto & \left\langle \ln F(\boldsymbol{t}, \boldsymbol{x}, \boldsymbol{\delta})+\ln p(\boldsymbol{x}|\tau)+\ln p(\textbf{copula}) \right\rangle_{\boldsymbol{\tau}}\nonumber\\
  &\propto& \left\langle \sum_{i=1}^{n}\left( \frac{z_{i}}{2}-\lambda(\delta_{i})z_{i}^{2}\right)-\frac{ \boldsymbol{x}^{2}}{2\boldsymbol{\tau}}-\frac{\boldsymbol{v}^{T}(\Sigma_{v}^{-1}-I)\boldsymbol{v}}{2}\right\rangle_{\boldsymbol{\tau},\boldsymbol{n}}\nonumber\\
&\propto& -\boldsymbol{x}^{T}\boldsymbol{A}^{T}\Lambda_{\delta}\boldsymbol{A}\boldsymbol{x}+\frac{1}{2}(2t-1)^{T}\boldsymbol{A}\boldsymbol{x}\nonumber\\
&& -\frac{1}{2}\left\langle \boldsymbol{x}^{T}\Lambda_{\tau}\boldsymbol{x}\right\rangle_{\boldsymbol{\tau}} -\frac{\boldsymbol{v}^{T}(\Sigma_{v}^{-1}-I)\boldsymbol{v}}{2},
\end{eqnarray}
where $\Lambda_{\tau}\triangleq \text{diag}(\tau_{1}, ..., \tau_{m})$ and \cite{li2015robust} denoted that $\Lambda_{\delta}\triangleq \text{diag}(\lambda(\delta_{1}), ..., \lambda(\delta_{n}))$. $\langle \cdot\rangle$ evaluates expectation with respect
to a random variable.

Obviously $q_{x}(x)$ has a Gaussian posterior PDF with the following parameters:

\begin{eqnarray}\label{21}
  \boldsymbol{\mu}_{x} &=& \boldsymbol{\Sigma}_{x} \boldsymbol{A}^{T}\left( \frac{1}{2}(2t-1)-2\Lambda_{\delta}\boldsymbol{\mu}_{n}\right),\\
  \boldsymbol{\Sigma}_{x} &=& \left( \Lambda_{\langle\tau\rangle}+2\boldsymbol{A}^{T}\Lambda_{\delta}\boldsymbol{A}+\frac{(\Sigma_{v}^{-1}-I)}{\tau}\right)^{-1}.
\end{eqnarray}

The third part of (21) captures the dependency of wavelet coefficients using proposed copula model.

 \textbf{2. Update of}  $q_{\tau}(\boldsymbol{\tau})$:
The posterior PDF of $q_{\tau}(\boldsymbol{\tau})$ can be derived by
\begin{eqnarray}\label{22}
  \ln q_{\tau}(\boldsymbol{\tau}) &\propto& \left\langle - \frac{1}{2}\ln\boldsymbol{\tau}-\frac{\boldsymbol{x}^{2}}{2\boldsymbol{\tau}}-\frac{\boldsymbol{v}^{T}(\Sigma_{v}^{-1}-I)\boldsymbol{v}}{2}-\frac{\boldsymbol{\lambda}^{2}}{2}\boldsymbol{\tau}\right\rangle_{\boldsymbol{x},\boldsymbol{\lambda}},\nonumber\\
\end{eqnarray}
where $\tau$ has the Generalized Inverse Gaussian (GIG) distribution with the following parameters:
\begin{equation}\label{23}
  \boldsymbol{\tau} \sim \text{GIG}\left( \frac{1}{2}, \langle \boldsymbol{x}^{2} \rangle+\boldsymbol{x}^{T}(\Sigma_{v}^{-1}-I)\boldsymbol{x}, \langle\boldsymbol{\lambda}^{2}\rangle\right).
\end{equation}
\textbf{3. Update of $q_{\lambda}(\boldsymbol{\lambda})$:}  In the same way, variational approximation of  yields
\begin{equation}\label{24}
  \ln q_{\lambda}(\boldsymbol{\lambda}) \propto \langle \ln p(\boldsymbol{\tau}|\lambda) + \ln p(\boldsymbol{\lambda})\rangle_{\boldsymbol{\tau}}.
\end{equation}
By setting $f=0$, the posterior PDF of $\boldsymbol{\lambda}$ becomes Rayleigh distribution.
\begin{equation}\label{25}
  \boldsymbol{\lambda} \sim \text{Rayleigh}(\boldsymbol{\lambda}|\frac{1}{\sqrt{\langle\boldsymbol{\tau}\rangle}}).
\end{equation}

The posterior PDF of $\boldsymbol{n}$, $\boldsymbol{\beta}$ and $\boldsymbol{\delta}$ can be found in \cite{li2015robust}. The overall 1b-CS recovery based on DGVC-MDL algorithm is summarized in Algorithm 1.

\begin{algorithm}
\label{DLDOA}
\SetKwData{Left}{left}
\SetKwData{This}{this}
\SetKwData{Up}{up}
\SetKwFunction{Union}{Union}
\SetKwFunction{FindCompress}{FindCompress}
\SetKwInOut{Input}{input}
\SetKwInOut{Output}{output}
\caption{The proposed DGVC-MDL algorithm for 1b-CS recovery}
\Input{\\
Array output $\mathbf{t} \in \mathbb{R}^{n\times 1}$ based on model (\ref{1})\\
Measurement matrix $\mathbf{A} \in \mathbb{R}^{n\times m}$\\
}
\Output{\\
wavelet coefficients estimation $\mathbf{x}\in \mathbb{R}^{m}$\\}
\textbf{Initialize} $\tau$, $\lambda$ and $\delta$ \\
\While{$\mathrm{iter} < \mathrm{maxiter}$}{
$\boldsymbol{\hat{\eta}}=\frac{M_{s}}{\sum_{i=1}^{M_{s}}(\frac{(\boldsymbol{\hat{f}}+1)|x_{i}|}{\boldsymbol{\hat{f}}+\boldsymbol{\hat{\eta}}|x_{i}|})}$\;
update $\boldsymbol{\mu}_{x}$ and $\boldsymbol{\Sigma}_{x}$ using (\ref{21}) and (21)\;
update $\mathbf{\tau}$ using (\ref{22})\;
update $\mathbf{\lambda}$ using (\ref{25})\;
}
$\widehat{\mathbf{x}}\gets \boldsymbol{\mu}_{x}/\mathrm{norm}(\boldsymbol{\mu}_{x})$;\\
\end{algorithm}

\section{Simulation Results}
\label{sec: Sim}
The performance of the proposed DGVC-MDL algorithm is investigated in this section. The measurement  matrix $\textbf{\textit{A}} \in \mathbb{R}^{n \times m}$ is randomly drawn from $\mathcal{N}(0,1)$ with i.i.d entries and $\textbf{\textit{A}}$'s columns are normalized to unit value. Our proposed algorithm is compared with the BIHT algorithm \cite{jacques2013robust}, the 1-BCS, R1-BCS \cite{li2015robust} and AOP \cite{yan2012robust}. Note that the BIHT algorithm has been also simulated. However, due to the small SNR of this method, the corresponding results are not represented here. 

The SNR is utilized to demonstrate the performance of the proposed DGVC-MDL algorithm over $10^{2}$ independent runs. 
Fig. 3 illustrates the reconstruction SNR versus the sampling rate where the initial values of $\tau$, $\lambda$ and $\delta$ are $10^{-8}$, $10^{-8}$ and $1$, respectively. The sampling rate which equals $\frac{n}{m}$ is the oversampling ratio; it varies from $2$ to $6$. We set $m=1024$ for input image of size $32\times32$ and we consider the wavelet coefficients at the second decomposition of the SAR image \text{e2\_981006\_18093\_2889} \footnote{ The test image is chosen from the Coastal Environmental Assessment Regional Activity Centre (CEARAC) database which can be found in http://cearac.poi.dvo.ru/en/db/} and Baboon. When the sampling rate equals 2, the SNR is near to R1-BCS and 1-BCS, however when the sampling rate increases the SNR improves up to 2.5 dB compared to R1-BCS, one-BCS and AOP.


\begin{figure}
\centering     
\subfigure[Baboon-SNR]{\label{fig:a}\includegraphics[width=90mm]{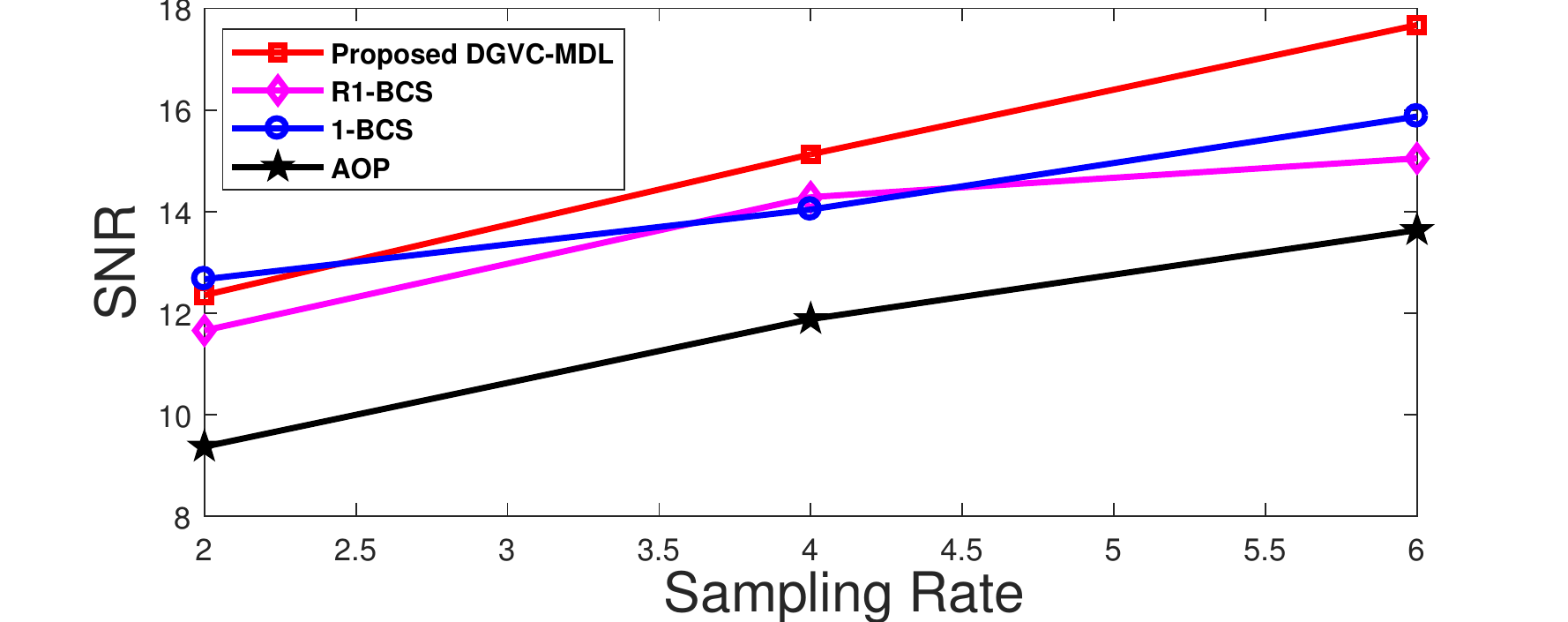}}
\centering
\subfigure[SAR-SNR]{\label{fig:b}\includegraphics[width=90mm]{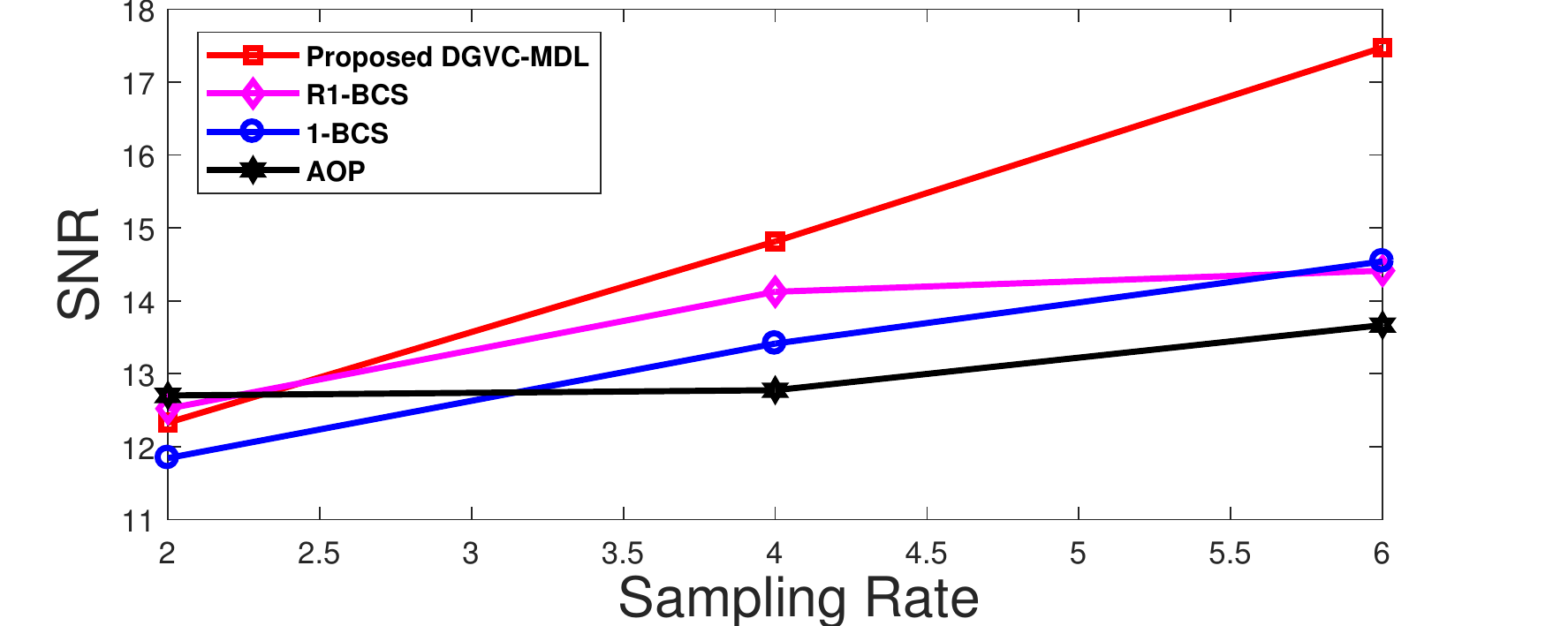}}
\caption{SNR versus sampling rate.}\label{Fig.2}
\end{figure}
\section{CONCLUSION}
\label{sec: con}
 In this paper, a new statistical model based on copula distribution is proposed for 1b-CS to capture the intrascale dependencies between wavelet coefficients. The VB of this new model is derived mathematically and it can estimate all unknown parameters in closed forms. The reconstruction SNR of the proposed algorithm is 2.5 dB better than SNR of other methods when the sampling rate is equal to 6.

%
\IEEEpeerreviewmaketitle

\ifCLASSOPTIONcaptionsoff
  \newpage
\fi

\newpage
\bibliographystyle{IEEEtran}

\bibliography{IEEEabrv,refrencesss}

\end{document}